\documentclass[manuscript]{aastex}
\usepackage{amsmath}
\usepackage{graphicx}
\usepackage{color}

\begin{document}

\shorttitle{X-Ray echo from {\it SWIFT} J1644+57}
\shortauthors{Cheng et al.}

\title{X-Ray afterglow of {\it SWIFT} J1644+57: a Compton echo?}
\author{K. S. Cheng$^{1}$, D. O. Chernyshov$^{1,2}$, V. A. Dogiel$^{1,2,3}$, Albert K. H. Kong$^{4}$, C. M. Ko$^{5}$}
\affil{$^1$Department of Physics, University of Hong Kong,
Pokfulam Road, Hong Kong, China}
\affil{$^2$I.E.Tamm Theoretical Physics Division of P.N.Lebedev
Institute of Physics, Leninskii pr. 53, 119991 Moscow, Russia}
\affil{$^3$Moscow Institute of
Physics and Technology, 141700 Moscow Region, Dolgoprudnii,
Russia}
\affil{$^4$Institute of Astronomy and Department of Physics, National Tsing Hua University, Hsinchu 30013, Taiwan}
\affil{$^5$Institute of Astronomy, Department of Physics and Center for Complex Systems, National Central University,
Jhongli, Taiwan}
\begin{abstract}
 {\it Swift},{\it Chandra} and {\it XMM} have found a weak but nearly constant X-ray component from {\it Swift} J1644+57  that appeared at $\sim$ 500 days and was visible at least until $\sim$ 1400 days after the stellar capture, which cannot be explained by standard tidal disruption theories. We suggest that this X-ray afterglow component may result from the Thomson scattering between the primary X-rays and its surrounding plasma, i.e. a Compton echo effect.  Similar phenomena has also been observed   from  molecular clouds in our Galactic Center, which were caused by the past activity of Srg A*. If this interpretation  of {\it Swift} J1644+57 afterglow is correct, this is the first Compton Echo effect observed in the cosmological distances.
\end{abstract}

\date{\today}

\maketitle

\section{Introduction}
Recent observations by X-ray and UV telescopes  have found flares from non-active galaxies,  that have been interpreted as    resulting from  star tidal disruption  by massive black holes in their nuclei  \citep[see the review of][and references therein]{komossa15}. These processes of tidal disruption are accompanied   by the release of a significant fraction of a solar rest mass of energy across the electromagnetic spectrum. These emissions show strong variability.

 Theoretical  models, e.g.  \citet{rees,phinney}, was analytically described as  thermal. The time-dependent brightness of the central source, $L(\bar t)$, was  nicely described as
\begin{equation}
L(\bar t) = L_0\left(\frac{\bar t}{t_0} + 1\right)^{-5/3}\,,
\end{equation}
where $\bar t$ is the current time,   and the parameters $L_0$ and $t_0$  are derived to reproduce the data.
 In addition to the thermal models above, non-thermal X-ray emission,
generated in relativistic outflows was also predicted by \citet{wong07}. Further studies by \citet{gian11} extended this by analogy with the GRB models of \citet{sari98}, deriving parameters for non-thermal radio emission. This non-thermal emission was apparently observed in {\it Swift} J1644+57 \citep{bloom},  and has also claimed in {\it Swift} J2058.4+0516 \citep{cenko} and {\it Swift} J1112.2-8238 \citep{brown15}, making these three events the proto-types of a class of relativistic tidal disruption flares.

The {\it Swift} J1644+57  transient X-ray source  was discovered in the direction of the constellation Draco \citep[see][]{burrows} with the peak luminosity $\sim 10^{48}$ erg s$^{-1}$. Observations  showed that the transient originated from the center of a galaxy at cosmological distances
 involving a massive black hole in the galaxy nucleus
\citep[see][]{levan,Yoon15}.  It was concluded that Sw J1644+57 was most likely originated from the central massive
black hole, and the X-ray emission  was interpreted as a result of stellar capture by a massive black hole \citep[see][]{burrows,levan}.

 Recently, {\it Swift} observations
 \citep{levan15}  have presented extremely unusual evolution of the X-ray
 lightcurve of {\it Swift} J1644+57. At the initial stage the luminosity decreased according the decay law $t^{-5/3}$  and for 500  days it decreased from $10^{-10}$   erg cm$^{-2}$ s$^{-1}$  to  $10^{-12}$ erg cm$^{-2}$ s$^{-1}$.

\begin{figure}
\plottwo{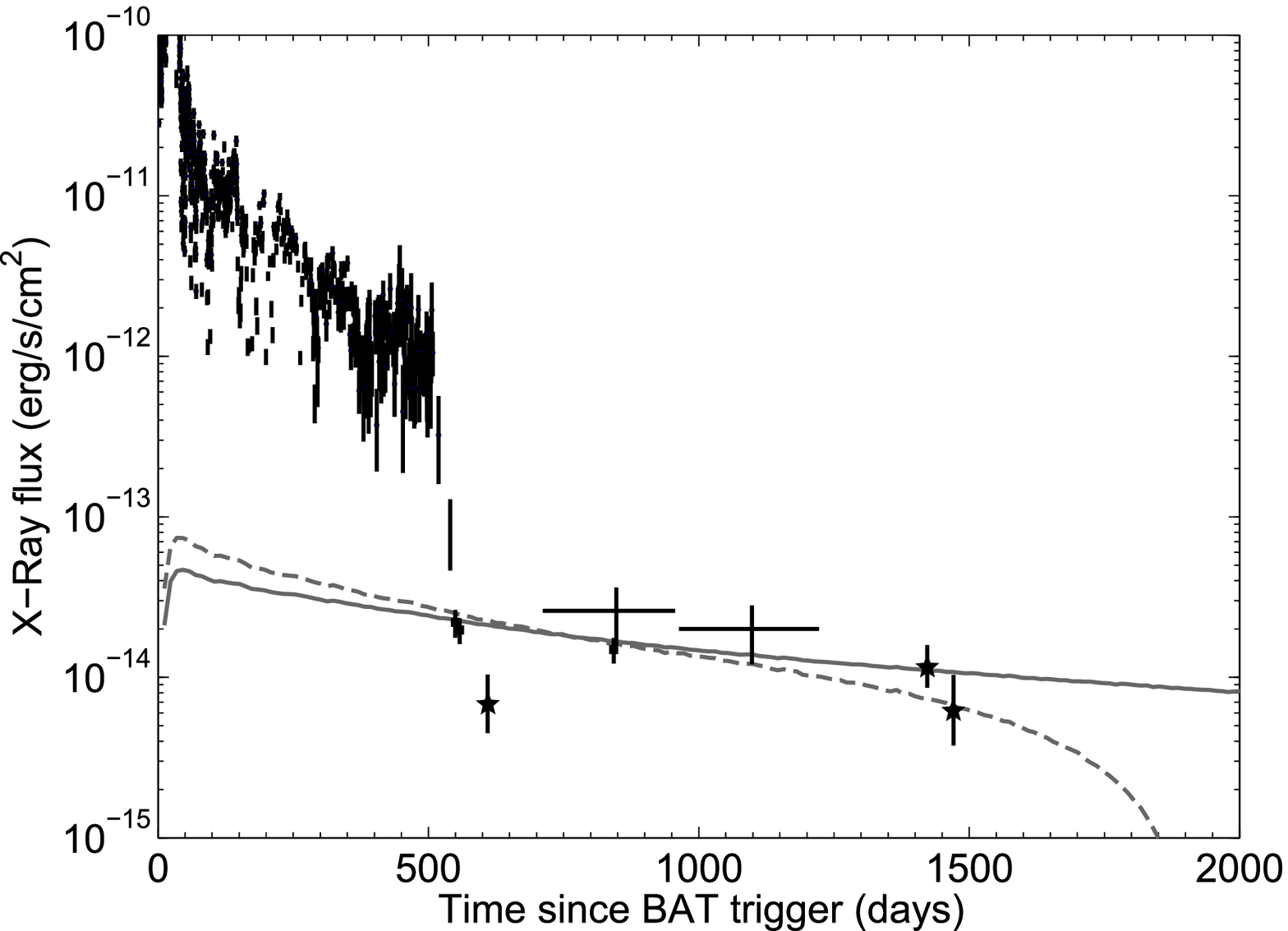}{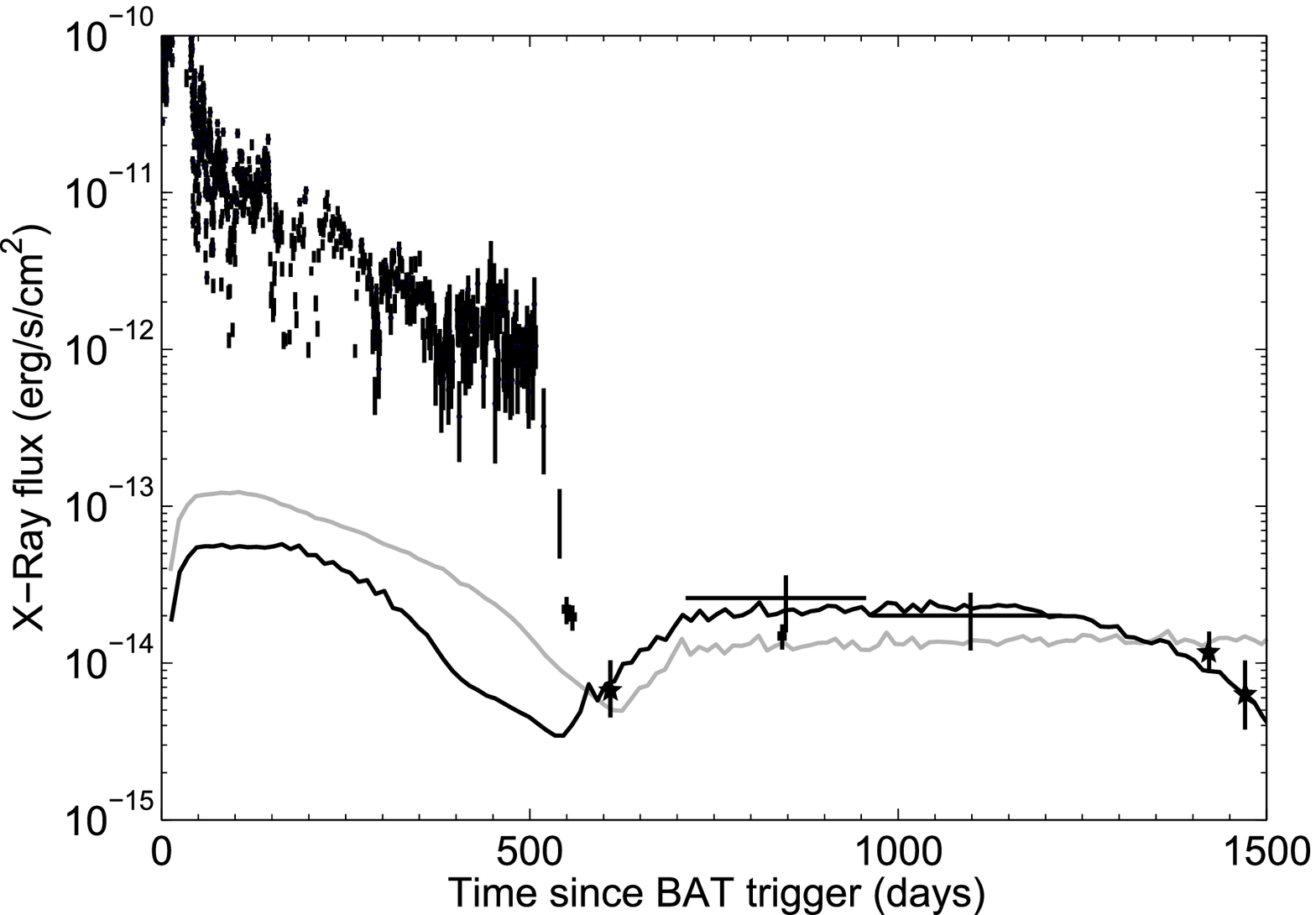}
\caption{   The observed light curve of  {\it Swift} J1644+57 and model fittings. The datapoints of {\it Swift}, {\it XMM-Newton} and  {\it Chandra} are shown by crosses, asterisks and squares, respectively. The model luminosity  of reflected emission (if generated by  Thomson scattering) is shown  for  two cases: Left panel: the solid line - density of spherically distributed ionised gas around the black hole $n_e=1500$ cm$^{-3}$ with the radius $R=2$ pc,   and the dashed line for  $n_e=3000$ cm$^{-3}$ and $R=0.8$ pc. Right panel: the spherical shell of background gas (see the text) for the internal radius of the shell $R_1=0.3$ pc and the external radius $R_2=0.65$ pc,  $n_e=10^4$ cm$^{-3}$, $\Delta\Omega=\pi$  (black curve), and $R_2=2$ pc, $n_e=10^4$ cm$^{-3}$, $\Delta\Omega=\pi/2$ (grey curve)}\label{fig1}.
\end{figure}

After 500 days  the source switched-off on a timescale less than ten days.  Beyond this time, for the next 1000 days, a weak X-ray component ($\sim 3\times 10^{-14}$   erg cm$^{-2}$ s$^{-1}$) was observed with an almost constant luminosity (see Fig. \ref{fig1}).
%\begin{figure}[h]
%\begin{center}
%\includegraphics[width=0.7\textwidth]{fig1.eps}
%\end{center}
%\caption{The observed light curve of  {\it Swift} J1644+57.  The model luminosity  of reflected emission (if generated by the Thomson scattering) is shown  for the two cases: the solid line - density of ionised gas around the black hole $n_e=1500$ cm$^{-3}$ with the radius $R=2$ pc,  the dashed-dotted line for $n_e=3000$ cm$^{-3}$ and $R=1$ pc, {\bf and the dashed line for  $n_e=3000$ cm$^{-3}$ and $R=0.8$ pc}. The datapoints of {\it Swift}, XMM-Newton and Chandra are shown by crosses, asterisks and squares, respectively }\label{fig1}
%\end{figure}

 It is also relevant to consider that  processes of energy release  in the form of X-ray flux near the central black hole occurred also in our Galaxy about hundred years ago although with a significantly smaller luminosity. From analyses of X-ray emission from the molecular clouds in the Galactic center, it was concluded that about 100 years ago the X-ray luminosity of our central black hole was several orders of magnitude higher ($\sim 10^{39}$ erg s$^{-1}$) \citep[see  the pioneer publications of][and subsequent publication of \citealt{yu11} and others]{suny,koyama} than at present ($ 10^{32}-10^{33}$ erg s$^{-1}$) \citep[see][]{ponti15}. This emission switched-off $\sim 100$ years ago, and is not seen now, but its afterglow is observed at present as:
\begin{itemize}
\item continuum X-ray emission due to Thomson scattering (reflection) of primary photons on  electrons of  molecular clouds \citep[][and others]{suny,terrier} and
\item a flux of the iron fluorescence K-$\alpha$ line at 6.4 keV which is generated by photo-ionization  of background iron atoms  \citep[see][and others]{koyama,ponti,clavel}.
\end{itemize}

As follows from the analysis  of the continuum and the 6.4 keV iron line emission in the direction of the cloud Sgr B2,  it was almost stationary for more than 5 years \citep[see e.g.][]{revn04}, which is about the characteristic time for X-ray photons to cross the cloud.  After that the intensity of emission from the cloud had dropped down, because the photon front left Sgr B2 \citep[][]{inui09,terrier,nobu11,nobu14,zhang15}.

 The cloud Sgr B2, whose  reflected emission is observed at present, is at a distance about 100 pc from the GC. It is reasonable to assume that processes of the Thomson scattering could take place nearby the black hole at the initial stage of the flare, because
there are many gaseous complexes in the close vicinity of the GC \citep[see][]{katia12}. The central black hole is surrounded by a warm ionized gas whose density is $910$ cm$^{-3}$ and the total mass is about 190 $M_\odot$. The radius of this region is about 1 pc.  There is also the Circumnuclear disk (CND) of molecular hydrogen at the distance beyond 1 pc from the GC. Its total mass is about $2\times 10^5M_\odot$, and the gas density is about $2\times 10^5$cm$^{-3}$.

 In this paper we speculate that  processes of  Thomson scattering may take place in the vicinity of the {\it Swift} J1644 center, which result in a weak but nearly constant luminosity X-ray afterglow. If the central source is surrounded by a gas, then it is inevitable that  a flux of reflected photons must be generated in this region. Since the observed weak X-ray component stays as constant at least up to 1400 days, then the size of reflection region must be $\sim 1$ pc.

\citet{levan15} have presented a detailed absorbed X-ray spectrum for {\it Swift} J1644+57. However, in order to calculate the reflected X-ray luminosity due to Thomson scattering, we need to use the unabsorbed X-ray luminosity as the primary injected X-rays. In section 2 we analyse the data of {\it Swift}, {\it Chrandra} and {\it XMM} to obtain the unabsorbed X-rays from {\it Swift} J1644. In section 3, we analyse the model of Thomson scattering for {\it Swift} J1644 and determine  parameters of background electrons needed to generate the observed weak X-ray component (afterglow). A brief discussion is presented in section 4.

\section{Light curve and spectrum of X-rays}
Swift J1644+57 has been monitored with the X-ray Telescope (XRT) onboard {\it Swift} in a regular basis since the outburst. Furthermore, {\it Chandra} and {\it XMM-Newton} each observed the source three times 500 days after the outburst. Here we  make use of all the archival X-ray observations of Swift J1644+57 to construct the long-term X-ray light curve. For the {\it Swift} data, we extracted the XRT light curve and energy spectra by using the XRT products generator \footnote{http://www.swift.ac.uk/user\_objects} \citep[][]{evans07,evans09}. We only used the photon-counting mode data in the 0.3--10 keV band. In general, every single observation results a data point, but for the last two data points, we have to combine several observations together in order to obtain sufficient signals.  We converted the XRT count rates into 0.3--10 keV unabsorbed flux using \texttt{PIMMS} by assuming an absorbed power-law spectrum with a photon index of 2 and absorption from our Galaxy \citep[$N_{H,Gal}=1.75\times10^{20}$ cm$^{-2}$;][]{will13}, and the host galaxy ($N_{H,host}=2\times10^{22}$ cm$^{-2}$ at $z=0.354$). The spectral parameters are based on spectral fits from early-time observations \citep[][]{burrows}. Figure 1 shows the long-term X-ray light curve of Swift J1644+57.

For {\it Chandra} data, all three observations were taken with the Advanced CCD Imaging Spectrometer array (ACIS-S). We used \texttt{CIAO} version 4.7 to process the data and only 0.3--7 keV photons were used to minimize the background. An extraction region of 2 arcsec in radius and a source-free background region were employed to extract the source counts. To convert the count rates into 0.3--10 keV absorbed fluxes with \texttt{PIMMS}, we assume the same spectral parameters applied to the XRT data.  Because there are too few photons (5--13) in each observation, we combined the three observations to obtain a rough spectrum. We fitted the spectrum with an absorbed power-law model and fixed the absorption at the values determined from the XRT data; we obtained a best-fit with a photon index of $1.4^{+2.0}_{-1.8}$ (90\% confidence).

We processed the {\it XMM-Newton} data with \texttt{SAS} version 14.0.0. Data in the energy range of 0.3--10 keV were extracted and source counts were obtained with a 15 arcsec radius region as well as a source-free background region. Like in the {\it Swift} and {\it Chandra} data, we convert the count rates into X-ray fluxes with \texttt{PIMMS} by assuming an absorbed power-law spectrum. We also combined the three observations together to get a time averaged spectrum in spite of low photon counts. Using the same simple power-law model and fixing the absorptions, we obtained a best-fit photon index of $1.9^{+1.5}_{-0.8}$, consistent with the fits obtained from {\it Swift} and {\it Chandra}.

 To benchmark our analysis we have compared our absorbed light curve with that given in \citet{levan15}. Our results are consistent with theirs.

\section{Parameters of the Model}

 The total isotropic energy emitted by  {\it Swift} J1644+57 for the first 20 days is about $3\times 10^{53}$ erg \citep[see][]{burrows}. That is most of the total energy  emitted for the whole time ($\sim 500$ days) of the flare.  

The density distribution of primary photons  can be presented as
\begin{equation}
n_{ph}(\rho,z,t) = \frac{L_x(t-r/c,\rho,z)}{4\pi r^2c}\exp(-\tau)\,,
\label{prim_density}
\end{equation}
where $\rho$ and $z$ are the cylindrical coordinates,  $r = \sqrt{\rho^2 + z^2}$.  The $z$-axis is directed from Earth to {\it Swift} J1644+57.  In this equation, we should take into account that the X-ray emission is not isotropic. The cone angle of the X-ray flux is defined as $\Delta\Omega$.  The luminosity $L_x$ is taken directly from the data shown in Fig. \ref{fig1}. We formally define an optical depth, $\tau$, traveled by a primary photon, if absorption in the medium is essential.

For estimates of the flux generated by the Compton echo, it does not matter what kind of background electrons surrounds the central source: whether they are electrons of a neutral hydrogen or electrons of fully ionised plasma.

 The density of background gas around Sw J1644+57 are very uncertain although \citet{berger12} derived model dependent estimate of the density from the observed radio emission of Sw J1644+57. These estimates are based on phenomenological analysis of radio emission provided, e.g. by \citet{sari98} and others for GRBs. We notice some essential differences between GRBs and processes of tidal disruption. The jet gamma factor in GRBs is $\Gamma>10^2$, whereas jets from tidal disruption are mildly relativistic with $\Gamma\ga 1-10$ and the beaming angle of the jet is also wider, $\sim 1/\Gamma$ \citep[see e.g.][]{liu15}.

As \citet{berger12} estimated the gas density around Sw J1644+57 is strongly nonuniform and decrease from 2 cm$^{-3}$ at the distance $r\sim 0.1$ pc  to 0.1  cm$^{-3}$at $r\sim 1$pc. Again we want to remark that the density deduced in \citet{berger12} is based on indirect method, which is strongly model dependent.

We present below the medium parameters around the central black hole in our Galaxy. As follows from \citet{katia12}, the Central Cavity within the radius $r\ga 1$ pc is filled by gases. The innermost of the Cavity  is filled with the hot ionised gas  with the temperature $T\sim 1.5\times 10^7$ and the density $n_H\sim 20$ cm$^{-3}$. These parameters of the hot gas are very similar to that derived by \citet{berger12}. But most of the Central Cavity is filled by the warm ionized gas \citep[$T \sim 7000$ K and $n_H\sim 10^3$ cm$^{-3}$, see][]{katia12}.

Because of these uncertainties of medium parameters at Sw J1644+57   we present below calculations for different cases of gas distribution in order to show a possibility that in principle  the Compton echo could be there. They are: a) a uniform gas distribution; b) a gaseous shell around Sw J1644+57 between the radiuses $R_1$ and $R_2$; and c) a dense molecular cloud nearby Sw J1644+57 of a size $r$.  In all cases the product $L_x n_e \Delta\Omega$   is chosen in the way that  the afterglow luminosity is at the level $\sim 3\times 10^{-14}$   erg cm$^{-2}$ s$^{-1}$.

If  photons of the afterglow emission are generated by the Compton echo  then the reflected photons have  the same delay time $t$ (in comparison with the time of direct propagation from the source to the observer) if they are scattered by electrons on a  surface, determined as \citep[see][]{suny98}
\begin{equation}
\frac{z}{c}=\frac{1}{2t}\left(t^2-\left(\frac{\rho}{c}\right)^2\right)\,.
\label{refl}
\end{equation}

\begin{figure}[h]
\begin{center}
\includegraphics[width=0.85\textwidth]{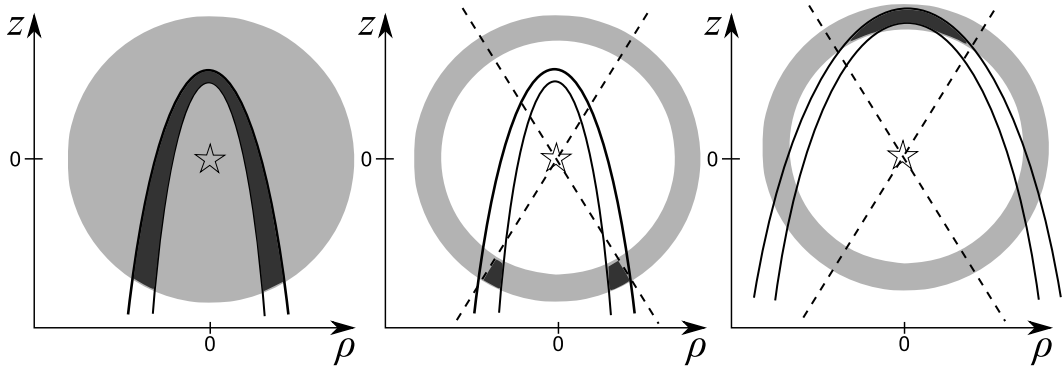}
\end{center}
\caption{ Schematic illustration for region between two parabolic lines (black shaded region) contributing to the Compton echo. Left panel: case for a spherical density gas (grey shaded region) and isotropic X-ray emission. Middle panel: case for gas with spherical shell distribution and X-rays emitted in a cone (within the dashed lines). In this case, the emission is generated by "side wings". Right panel: The same as in the Middle panel but for the case when the top of the delay region reaches the shell.}\label{fig2}
\end{figure}

If the source of primary photons is active for some period  e.g. $\Delta T$, then a flux of reflected photons is determined from the conditions of equal arrival time of delayed photons to the observer.  We define $T_1$ and $T_2$  as the time of the  beginning and   end of the flare, i.e $T_1=T_2+\Delta T$. The  observer detects photons with the delay time in the range $T_2\leq t\leq T_1$. The area of reflected photons is determined by Eq. (\ref{refl}) and is enclosed between the two parabolas calculated for $t=T_1$ and $t=T_2$, see Figs. \ref{fig2}.

 The emissivity  of the reflected X-rays can be calculated as
\begin{equation}
\epsilon(\rho,z,t) = n_{ph}(\rho,z,t)n_e(\rho,z)\sigma_T(\Theta)\cdot\exp(-\tau_R)\,,
\label{sec_density}
\end{equation}
where $n_{ph}$ is defined by Eq. (\ref{prim_density}). In our case the optical depth , $\tau_R\ll 1$. The cross-section of scattering in the angle $\Theta$ is
\begin{equation}
\sigma_T(\Theta) = 4\pi r_e^2\frac{1 + \cos^2 \Theta}{2}\,,
\end{equation}
where $r_e$ is the classical electron radius and the reflection angle is defined as $\cot\Theta = z/r$. The distribution of background electrons is denoted as $n_e(\rho,z)$.

In order to estimate the luminosity of reflected photons at each moment, $t$, we should integrate the emissivity, $\epsilon(\rho,z,t)$ over the region filled by photons with the same delay time
\begin{equation}
L_x(t)=2\pi n_ec \iint\limits_S dz \rho d\rho\sigma_T (\Theta)n_{ph}(\rho,z,t)\,,
%\frac{\Delta\Omega}{4\pi}
\end{equation}
where the area of integration $S$ for the case of uniform gas distribution is enclosed between those three curves shown in  Fig. \ref{fig2}, i.e.
\begin{eqnarray}
&&\rho=cT_1\sqrt{1-\frac{2z}{cT_1}}\,,\\
&&\rho=cT_2\sqrt{1-\frac{2z}{cT_2}}\,,\\
&&\rho=\sqrt{R^2-z^2}\,,
\end{eqnarray}
within the cone of X-rays.

  The resulting  afterglow emission for the simplest case of uniform distribution are shown in Fig. \ref{fig1} (Left panel). The solid line shows the case of electron distribution with radius  $R= 2$ pc and the density of electrons  $n_e=1500$ cm$^{-3}$. The dashed line shows the case for $R=0.8$ pc and $n_e=3000$ cm$^{-3}$.

As one can see from the figure, the datapoints show a trend of slow decrease of the afterglow luminosity with time, our model describes the datapoints reasonably well.
It should be noticed that the afterglow luminosity  drops down sharply when the reflection region has crossed over the volume filled by the background gas, i.e., for the time $t>R/c$.

 The temporal variations of the case where electrons are distributed uniformly in the shell is shown in Right panel of Fig. \ref{fig1}. The black line is the case of the shell between $0.3 - 0.65$ pc, and $n_e=10^4$ cm$^{-3}$ and $\Delta\Omega=\pi$, while the gray line is for the case of the shell between $0.3 - 2$ pc, and $n_e=10^4$ cm$^{-3}$ and $\Delta\Omega=1$. The two peaks appear in the Right panel of Fig. \ref{fig1} can be explained by the illustration in  Fig. \ref{fig2}. The first peak corresponds to the time when the top of delay region does not reach the shell (Middle panel of  Fig. \ref{fig2}), and the second peak corresponds to the time when the top of delay region  reaches the shell (Right panel of  Fig. \ref{fig2}).

\begin{figure}[h]
\begin{center}
\includegraphics[width=0.5\textwidth]{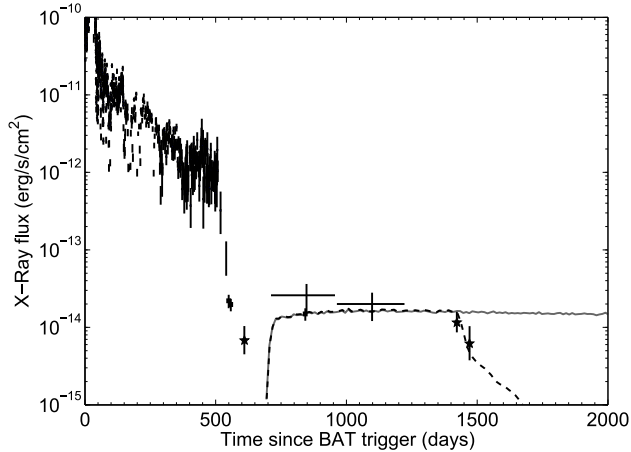}
\end{center}
\caption{ The temporal variations of the afterglow emission from a cloud which is at a distance 0.3 pc from Sw J1644+57. The gas density there is about $n_e=10^5$ cm$^{-3}$, $\Delta\Omega=0.2$ and the cloud size is 0.3 pc (the dashed curve) and 2 pc (solid curve).
}\label{fig3}
\end{figure}

In Fig. \ref{fig3} we showed the case in which the Compton echo is produced by a  cloud located  at the distance 0.3 pc from the black hole but in the opposite direction of radio jet.

\section{Discussion and Conclusion}
In our analysis we show that the quasi-stationary afterglow, detected  in the direction of {\it Swift} J1644+57, can be due to the Thomson scattering (Compton echo) of primary X-ray photons of the transient source on background electrons. This process is naturally expected if the transient source is surrounded by a background gas. A similar process was observed in our Galaxy where the flux of reflected photons, emitted by the central supermassive black hole 100 years ago, was detected in the direction of dense molecular  clouds. For the case of {\it Swift} J1644+57 the luminosity of the afterglow can be provided by processes of  Thomson scattering if the background gas density is high enough. Our calculations showed that the required value of the afterglow luminosity could be obtained if the gas density is  $> 1000$ cm$^{-3}$. We notice that the supermassive black hole in our Galaxy is surrounded by a warm ionized gas with the density about 1000 cm$^{-3}$.

 However, if we accept the  parameters of  background gas, as derived by \citet{berger12} from the radio data, then the only model, which can survive out of the three proposed models, is the third one, in which a high density cloud is located in the opposite direction of the radio jet. Therefore,  our model does not conflict with the parameters derived by \citet{berger12}.  

The region of reflection in the background gas is determined by
the condition when photons emitted at different times come after the Thomson scattering to the observer at the same time.
 For a prolonged flares the region of reflection is determined by the condition of  equal delay time for each moment of the flare.

For the case of {\it Swift} J1644+57 most of the transient energy is emitted during first $\sim 10$ days. Therefore characteristics of the afterglow (intensity, spectral index etc.)  reflect  parameters of the first stage of the transient.   The spectral index of the {\it Swift} J1644+57  transient at the initial stage is about -2 for the 1 to 10 keV range \citep[see][]{bloom,burrows,cheng12}. Then, the spectral index of the afterglow,  is expected also to be -2 for the Thomson process. However, poor statistics of the afterglow photons does not allow to derive a reliable estimate of this value.

 Our model predicted X-ray afterglow light curve does not sensitively depend on the exact form of the primary X-ray light curve if most of the primary X-rays are injected in a short time scale in comparing with propagation time in the plasma. The model afterglow  light curve shows a very slow decrease as the region of reflection propagates through the background gas. When the primary X-rays reaches the border of gas distribution, the afterglow luminosity switches-off rapidly.

In the conclusion we notice, that if our interpretation is correct, then the  {\it Swift} J1644+57 afterglow would be the first case of the Compton echo at cosmological distances.

 A direct confirmation of our model would be the detection of the iron fluorescent K-$\alpha$ line in the direction of {\it Swift} J1644+57, which should be generated by photoionization of background iron atoms. Our estimates showed that for the case of  {\it Swift} J1644+57 the line intensity is only $\sim 10^{-8}$ photons cm$^{-2}$s$^{-1}$ which cannot be detected by X-ray telescopes at present.

\section*{Acknowledgements}

 We would like to thank the referee for his very important suggestions, which significantly improve the content of our paper. The authors are very grateful to Katia Ferri\`{e}re for her very useful comments about conditions near Sw J1644+57 and the Galactic center.
KSC is supported by the GRF Grants of the Government of the Hong Kong SAR under HKU 701013. DOC and VAD  acknowledge a partial
support from the RFFI grants 15-52-52004, 15-02-02358. DOC is supported in parts by Dynasty Foundation.  KSC, DOC, VAD and AKHK  acknowledge support from the International Space Science Institute-Beijing to the International Team "New Approach to Active Processes in Central Regions of Galaxies". CMK is
supported, in part, by the Taiwan Ministry of Science and
Technology grants MOST 102-2112-M-008-019-MY3 and
MOST 104-2923-M-008-001-MY3.

\end{document}